\documentclass[12pt,letter]{revtex4}
\pdfoutput=1 

\usepackage{amsfonts}
\usepackage{dsfont}
\usepackage{amsxtra}
\usepackage{hyperref}

\usepackage{graphicx,epsfig}
\usepackage{dcolumn}
\usepackage{bm}

\newcounter{theappend}

\newcommand{\be}{\begin{equation}}
\newcommand{\ee}{\end{equation}}
\newcommand{\bea}{\begin{eqnarray}}
\newcommand{\eea}{\end{eqnarray}}
\newcommand{\beaa}{\begin{eqnarray*}}
\newcommand{\eeaa}{\end{eqnarray*}}

\newcommand{\ba}{\begin{array}}
\newcommand{\ea}{\end{array}}
\newcommand{\bi}{\begin{itemize}}
\newcommand{\ei}{\end{itemize}}
\newcommand{\ben}{\begin{enumerate}}
\newcommand{\een}{\end{enumerate}}

\newcommand{\td}{\tilde}

\newcommand{\lb}{\label}
\newcommand{\g}{\gamma}

\newcommand{\dl}{\delta}

\begin{document}

\title{On discrepancy between ATIC and Fermi data
\bigskip
\bigskip}

\author{Dmitry Malyshev}
 \email{dm137@nyu.edu}
 \altaffiliation{On leave of absence from ITEP, Moscow, Russia, B. Cheremushkinskaya 25}
\affiliation{ CCPP, 4 Washington Place, Meyer Hall of Physics, NYU, New York, NY 10003}

\date{\today}

\begin{abstract}
\noindent 
\bigskip
\bigskip
\bigskip

Either ATIC or Fermi-LAT data can be fitted 
together with the PAMELA data by three components:
primary background $\sim E^{-3.3}$, secondary background $\sim E^{-3.6}$, and an 
additional source of electrons $\sim E^{-\g_a} {\rm Exp}(- E / E_{\rm cut})$.
We find that the best fits for ATIC + PAMELA and for Fermi + PAMELA
are approximately the same, $\g_a \approx 2$ and $E_{\rm cut} \sim 500$ GeV.
However, the ATIC data have a narrow bump between 300 GeV and 600 GeV
which contradicts the smooth Fermi spectrum.
An interpretation of the ATIC bump as well as the featureless Fermi spectrum
in terms of dark matter models and pulsars is discussed.

\end{abstract}


\maketitle

\newpage



The question of interpretation of Fermi-LAT \cite{Abdo:2009zk},
HESS \cite{Collaboration:2008aaa,Aharonian:2009ah}, 
and ATIC data \cite{:2008zzr}
can be split into two parts: general properties of the flux and the presence
of features.
If one takes into account PAMELA data \cite{Adriani:2008zr}, 
then both Fermi and ATIC require
the existence of an additional flux of electrons and positrons complementary
to the standard primary and secondary backgrounds.

We will consider the following form for the additional flux
\be
\lb{addl_flux}
F_a \sim E^{-\g_a} e^{- \frac{E}{E_{cut}}}.
\ee
The general properties of the flux will be parameterized by 
the index $\g_a$ and the exponential cutoff $E_{cut}$.
Significant deviations from this form will be considered as ``features".

We will assume the primary background $\sim E^{- \g_{p}}$, 
the index $\g_{p} \approx 3.3$ can be estimated from 
the electron injection index $2-2.5$ due to shock acceleration
in the supernovae explosions \cite{1991ApJEllison,1992ApJReinolds}
(a similar estimation for the shock acceleration
in gamma ray bursts can be found in 
\cite{Achterberg:2001rx,Piran:2004ba}).
The shift of the index from $2-2.5$ to $3.3$ is due to cooling during propagation \cite{Longair1992}.
Secondary background
of electrons and positrons is produced by collisions of 
high energy protons and nuclei 
$\sim E^{-2.7}$ \cite{Strong:2007nh,Amsler:2008zzb} with the dust.
We will assume the secondary background
$\sim E^{- \g_{s}}$, $\g_{s} \approx 3.6$ where the difference in the indices
is again due to cooling.
Since the normalization and the indices of the backgrounds are known only
approximately, we will treat them as independent parameters in the fits.

\begin{figure}[t] 
\begin{center}
\epsfig{figure = 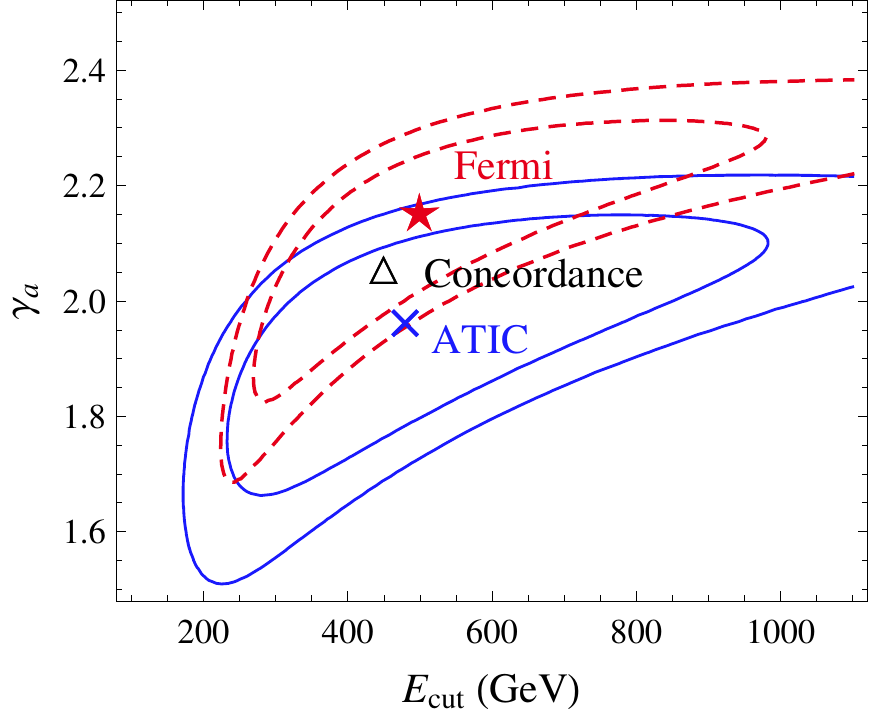,scale=0.9}
\end{center}
\vspace{-8mm}
\noindent
\caption{\small The results of fitting
the primary and secondary backgrounds together with 
an additional source of electrons and positrons 
to Fermi + PAMELA 
and to  ATIC + PAMELA. 
As everywhere else in the paper, for PAMELA, only $E > 10$ GeV points are used.
The best fit values and the parameter ranges are given in Table
\ref{FitData}.
The star (cross) represents the best fit to 
Fermi + PAMELA (ATIC + PAMELA)
with the reduced chi-squared $\chi^2_{\rm r\; best} = 0.4$ 
($\chi^2_{\rm r\; best} = 1.5$).
The dashed (solid)
contours correspond to $\chi^2_{\rm r\; best} + 1$ and $\chi^2_{\rm r\; best} + 2$
for Fermi (ATIC) + PAMELA.
The concordance model has
$\chi_{\rm r}^2 = 0.9\:(2.0)$ for Fermi (ATIC) + PAMELA.
}
\label{ConcordContour}
\vspace{1mm}
\end{figure}

\begin{figure}[t] 
\begin{center}
\epsfig{figure = 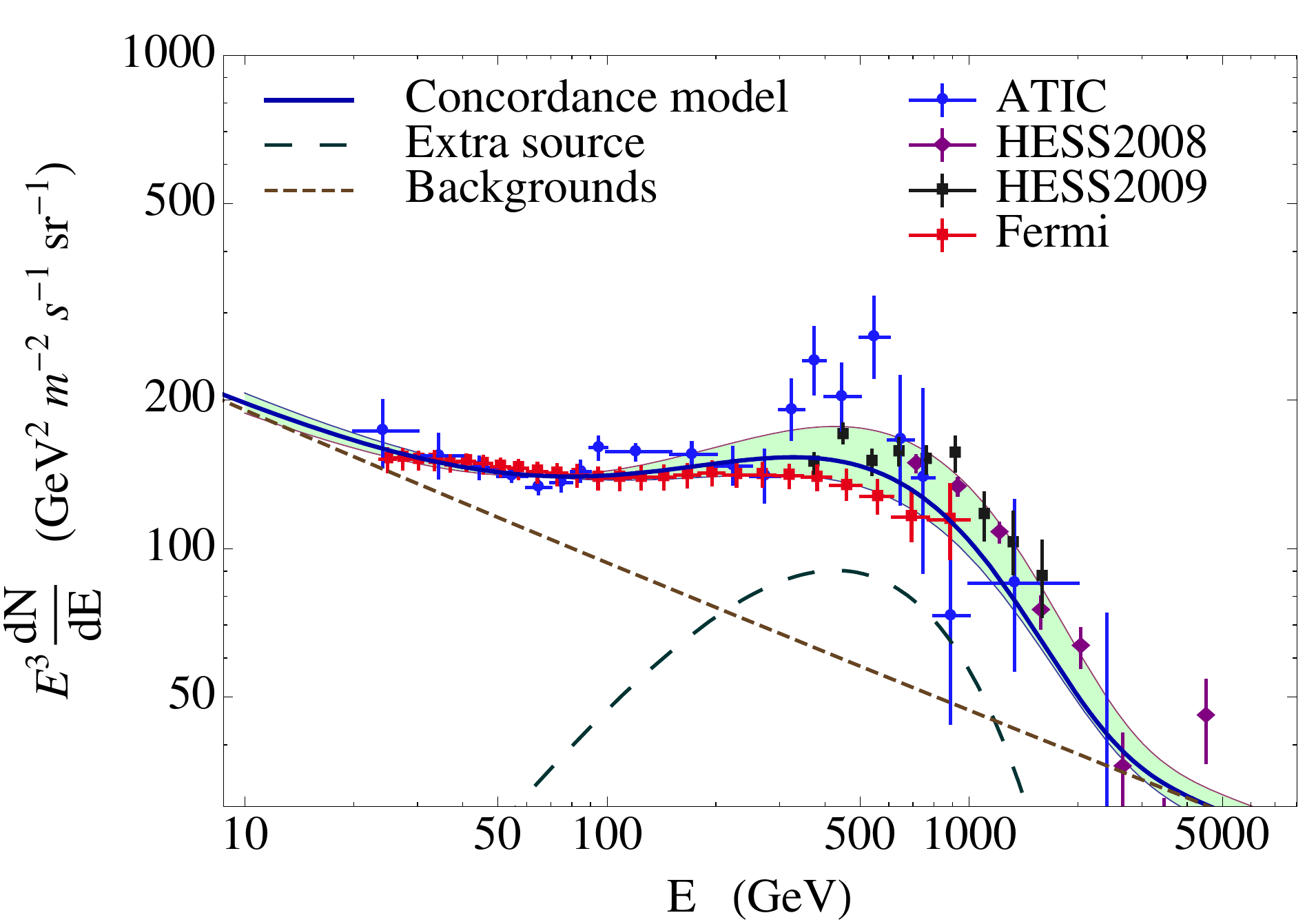, scale=0.5}\\
\epsfig{figure = 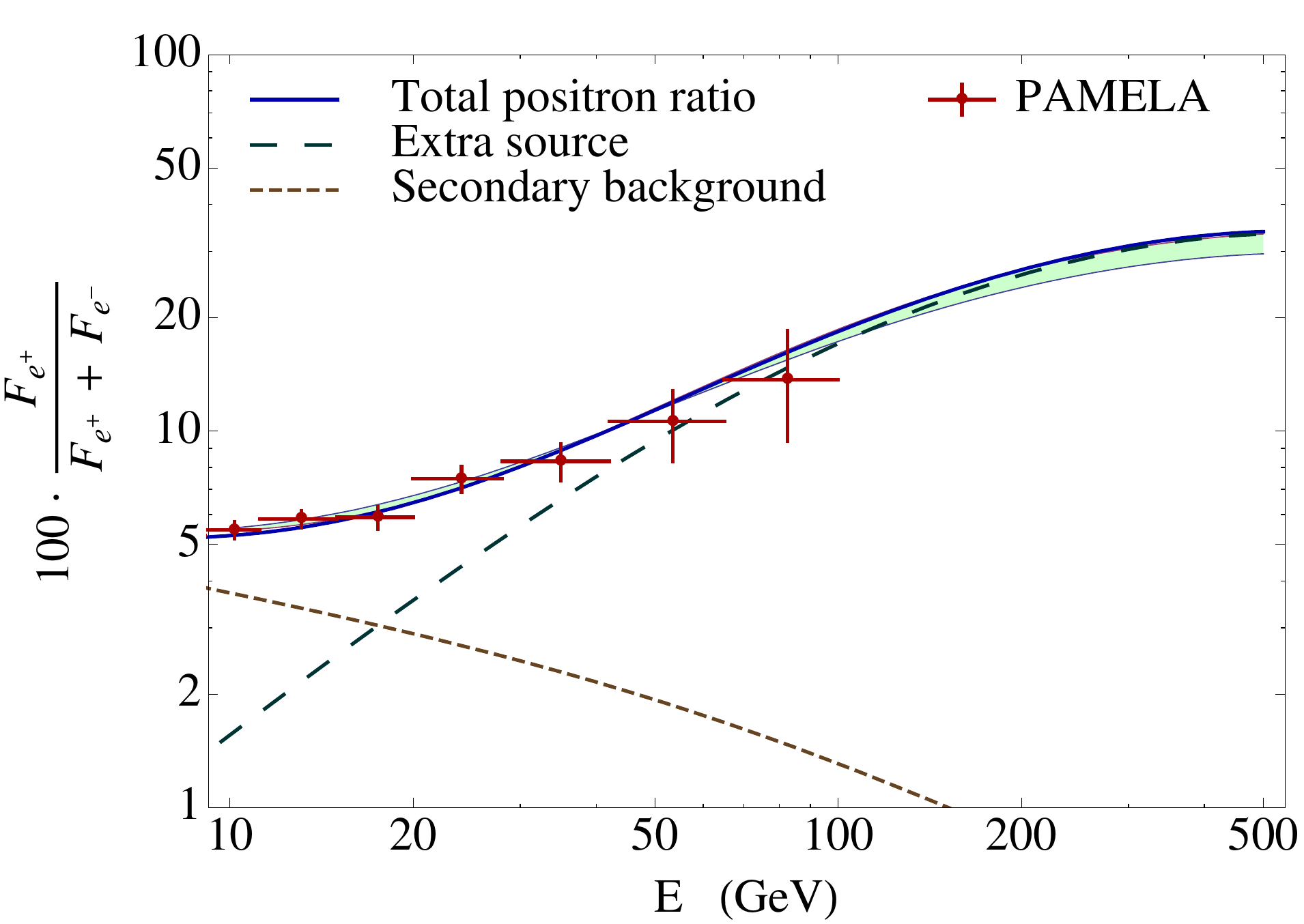, scale=0.5}
\end{center}
\vspace{-2mm}
\noindent
\caption{\small The concordance model corresponding to the triangle in Figure 
\ref{ConcordContour} with the parameters given in Table \ref{FitData}.
The upper (lower) edge of the band is the best fit for ATIC (Fermi) + 
PAMELA points with energies $E > 10$ GeV.
HESS 2008 and HESS 2009 points fit the concordance model, if multiplied
by a factor $\sim 0.9$ which may be related to contamination by diffuse gamma rays.
}
\label{ConcordPlots}
\vspace{2mm}
\end{figure}

In order to find the properties of the additional flux,
we will use the ATIC data, the Fermi data, and
the PAMELA data above 10 GeV
(the points below 10 GeV are assumed to suffer from solar modulation 
and we will discard them for the purposes of current analysis).
The results of the fits
to ATIC + PAMELA and 
to Fermi + PAMELA are presented in Figs.
\ref{ConcordContour} and \ref{ConcordPlots} and in Table \ref{FitData}.
Both ATIC + PAMELA and Fermi + PAMELA are fitted best by an additional
flux with $\g_a \approx 2$ and $E_{\rm cut} \sim 500$ GeV,
i.e., these experiments are consistent with each other from the point of view 
of general properties of the flux parameterized by Eq.~(\ref{addl_flux}).
It should be noted that without PAMELA data, the ATIC bump is better fitted with 
a harder additional flux, $\g_a \approx 1.7$, while the Fermi data are consistent 
with the primary background $\g_p \approx 3.0 - 3.1$ \cite{Abdo:2009zk, Grasso:2009ma}.

The second question is the presence of features.
The Fermi data are well approximated by
the additional source in Eq.~(\ref{addl_flux})
while the ATIC data have a deviation from (\ref{addl_flux})
between 300 GeV and 600 Gev.
Thus, although the parameters of the additional flux in Eq.~(\ref{addl_flux})
necessary to fit ATIC and Fermi data are similar,
the ATIC data have a bump while the Fermi data don't have any features,
i.e., there is a disagreement between the two data sets
at energies 300 - 600 GeV.
As one can see in Table \ref{FitData}, the best fit for Fermi + PAMELA
has a better reduced chi-squared,
$\chi_{\rm r}^2 = 0.4$,
than the best fit for ATIC + PAMELA,
$\chi_{\rm r}^2 = 1.5$.

In the following we will summarize the properties of dark matter (DM) models and pulsars
necessary to reproduce both the flux with features and the flux without features.
Due to energy losses, the source of high energy electrons should be close to 
Earth, within approximately 1 - 3 kpc.
For the purposes of flux calculation, the DM distribution can be viewed as homogeneous and
constant \cite{Kuhlen:2009is}
(unless there is a significant contribution from a local DM substructure
such as a clump \cite{Hooper:2008kv,Brun:2009aj,Kuhlen:2009is}).
The flux from a homogeneous source is \cite{Longair1992}
\be
\lb{FDM}
F_{\rm DM} = \frac{c}{4\pi}\frac{1}{b(E)} \int_{E}^{M_{\rm DM}} Q_{\rm DM}(\td E) d\td{E},
\ee
where $\dot E \equiv - b(E)$ is the energy losses.
At $E > 10$ GeV the energy losses are due to Inverse Compton Scattering and 
synchrotron radiation in the galactic magnetic field, thus $b(E) = b_0 E^2$ \cite{Longair1992}.
$Q_{\rm DM}$ is the source function for $e^+e^-$ produced by 
annihilating or decaying DM, $Q_{\rm DM} = \frac{dN}{dEdVdt}$.

\begin{table}[h]
\begin{center}
\begin{tabular}{|*{7}{@{\hspace{2.5mm}}c@{\hspace{2.5mm}}|}}
\hline
Data	&	$\g_a$	&	$E_{\rm cut}$	&	$E^3_0F(E_0)$      &      $\chi_{\rm r}^2$ & 
$\g_{\rm p}$ & $\g_{\rm s}$ \\
&	&	(GeV)	&	($\text{GeV}^2 \text{m}^{-2} \text{s}^{-1} \text{sr}^{-1}$)   & & & \\
\hline\hline
ATIC + PAMELA 	& $1.95 \pm 0.15$ & $480 \pm 200$ & $50 \pm 8$ & $1.5 - 2.5$ &
3.27 & 3.63 \\
\hline
Fermi + PAMELA & $2.15 \pm 0.10$ & $500 \pm 150$ & $44 \pm 6$ & $0.4 - 1.4$ &
3.31 & 3.63 \\
\hline
Concordance model & $2.05$ & $450$ & $47$ & 1.3 &
3.29 & 3.63 \\
\hline\hline
\end{tabular}
\end{center}
\vspace{-1mm}
\caption{\small
Numerical values for the fits presented in Fig.~\ref{ConcordContour}.
The normalization is given for $E_0 = 100$ GeV.
The ranges of parameters correspond to the ranges of reduced chi-squared
$\chi^2_{\rm r}$ in the fifth column.
The error bars for Fermi are computed as square root of systematic plus statistical errors squared.
$\g_p$ and $\g_s$ are the indices of the primary and secondary backgrounds
respectively.
The $\chi_{\rm r}^2$ for the concordance model is computed using Fermi,  ATIC,
and PAMELA ($> $10 GeV) points.
The best fits are found by varying 7 parameters:
2 indices and 2 normalization constants for primary and secondary backgrounds 
together with the index, the normalization and the cutoff of
the additional flux in Eq.~(\ref{addl_flux}).
}
\label{FitData}
\end{table}

The cutoff energy $E_{\rm cut}$ is the energy where the integral on the right hand side of 
Eq.~(\ref{FDM}) is saturated.
For energies $E \ll E_{\rm cut}$ the integral is insensitive to the variations of the 
lower limit and $F_{\rm DM} \sim 1 / b(E)$.
Thus an index $\g_a \approx 2$ is a universal prediction of DM.
Local clumps of DM result in a harder spectrum $\g_a < 2$
\cite{Hooper:2008kv,Kuhlen:2009is}. 
The dependence on the host halo profile leads to a softer spectrum $\g_a > 2$
at low energies \cite{Kuhlen:2009is}.
In models with small diffusion height, there may be some hardening $\g_a < 2$ at
low energies due to leakage of electrons from the Galaxy (see, e.g., 
model M2 in Figure 3 of \cite{Chen:2008fx}).

For a given $E_{\rm cut} \sim 500$ GeV, the DM mass and the shape of the spectrum
depend on the DM model.
In general, DM annihilation is followed by a sequence of decays leading to electrons 
and positrons together with other stable particles in the end.
Models with many steps in the decay process have smooth $e^+e^-$ spectrum and 
a large DM mass $M_{\rm DM} \gg E_{\rm cut}$ \cite{Kuhlen:2009is,Mardon:2009rc}.
These models are favored \cite{Meade:2009iu,Bergstrom:2009fa} 
by featureless Fermi spectrum.
Models with decay channels through $W$, $Z$ gauge bosons and quarks
have stricter constraints due to absence of
significant deviations from the expected backgrounds for
anti-protons \cite{Adriani:2008zq}
and diffuse gamma rays from the Galactic center 
\cite{Aharonian:2006au,Aharonian:2006wh}.
The flux of gamma rays from DM clumps assuming $b\bar b$ annihilation channel 
was also estimated \cite{Kuhlen:2008aw}.
DM models with small DM mass $M_{\rm DM} \lesssim 500$ GeV
(e.g., \cite{Bergstrom:2008gr,Grajek:2008pg})
seem to be in tension with Fermi data due to absence of significant
step-like features below or around 500 Gev.

DM models with few decay steps have a sharper cutoff
and $M_{\rm DM} \gtrsim E_{\rm cut}$ 
\cite{Kuhlen:2009is,Mardon:2009rc,Meade:2009rb}.
These models give reasonable fits to the Fermi data 
for $M_{\rm DM} \lesssim 1$ TeV \cite{Grasso:2009ma},
however, due to a sharp cutoff near $M_{\rm DM}$ they, generally,
fit better the ATIC data \cite{Mardon:2009rc}.
These models may have additional constrained due to final state radiation from 
the Galactic center and the Galactic ridge \cite{Mardon:2009rc,Meade:2009rb}.

Local clumps may produce additional features at high energies.
The presence of a large local clump is disfavored by Fermi but consistent with ATIC
\cite{Kuhlen:2009is}.
Furthermore, in models with many decay steps,
a significant contribution from a local clump may be necessary to fit the ATIC bump.

Thus, the Fermi data favor dark matter with many decay steps and a large
DM mass while ATIC requires either a DM model with few decay steps or a significant
local substructure in DM density distribution.

Let us now turn to pulsars.
In the calculation of electron and positron fluxes, pulsars can be considered
as point-like instantaneous sources \cite{Malyshev:2009tw}, 
$Q \sim \dl(x - x_0)\dl(t - t_0)$.
The main reason is that the typical propagation time ($\gtrsim 100$ kyr)
is much larger than the characteristic time scale when a pulsar loses most of its
rotational energy and the electrons and positrons are released to the interstellar medium
(ISM).

For every pulsar, we will consider two energy scales 
\cite{Malyshev:2009tw,Grasso:2009ma}:
the cutoff in the injection spectrum of electrons and positrons from the pulsar into 
the ISM, $E_{\rm inj.cut}$, which can be between few hundred
GeVs and tens of TeVs (see, e.g., \cite{Arons1996}
and references in \cite{Malyshev:2009tw}),
and the cooling break $E_{\rm br} = \frac{1}{b_0 t}$, obtained by integrating the 
energy losses $\dot E = - b_0 E^2$ during the propagation time $t$, where $t$ is 
approximately the age of the pulsar.
At $E \approx 100$ GeV the energy losses can be estimated as 
$b_0 \approx 1.6 \text{GeV}^{-1}\text{s}^{-1} \approx 5 \text{TeV}^{-1}\text{Myr}^{-1}$
\cite{Kobayashi:2003kp}.
At energies $E \gtrsim 100$ GeV the coefficient $b_0$ 
slightly decreases with the energy \cite{Kobayashi:2003kp},
since the Thompson approximation to inverse Compton scattering
between electrons and the starlight becomes inapplicable.

For pulsars with age $t \lesssim 10$ Myr,
the cooling break is $E_{\rm br} \gtrsim 20$ GeV.
In the ATNF catalog there are several hundred pulsars within 3 kpc from 
Earth and an age $t < 10$ Myr \cite{Manchester:2004bp}.
Below approximately 300 GeV,
the corresponding flux is well
approximated by the flux from a continuous distribution of pulsars in the Galactic plane
\cite{Malyshev:2009tw}.
An index $\g_a \approx 2$ requires an index in the
electron injection spectrum from pulsars
$\g_{\rm inj} \lesssim 2$ 
\cite{Hooper:2008kg,Yuksel:2008rf,Profumo:2008ms,Ioka:2008cv,Shaviv:2009bu,
Malyshev:2009tw,Kawanaka:2009dk}.

At energies $E \gtrsim 300$ GeV,
the flux receives contributions only from 
young pulsars $t \lesssim 1$ Myr within a smaller distance
$d \lesssim 1$ kpc (since the propagation time is smaller).
One may expect only of order ten such pulsars \cite{Manchester:2004bp}.
There is also a lower bound on the age of pulsars, 
a few tens of kyr, due to the fact that the electrons may still
be trapped by the Pulsar Wind Nebulae.

Depending on the relative value of $E_{\rm inj.cut}$ and $E_{\rm cut}$,
there are two possibilities for the $e^+e^-$ flux from pulsars:
\bi
\item $E_{\rm inj.cut} \sim E_{\rm cut}$, 
then the observed spectrum is naturally flat if we assume that the injection spectrum 
from pulsars is flat.
This possibility is favored by the Fermi data \cite{Grasso:2009ma}.

\item $E_{\rm inj.cut} \gg E_{\rm cut}$, then the cutoff in the observed spectrum
is due to the cooling break which is much sharper than an exponential cutoff.
One should also expect a series of steps due to consecutive cooling break
cutoffs from different pulsars \cite{Malyshev:2009tw}.
This possibility is consistent with ATIC but may be in tension with Fermi.
\ei

To summarize, both Fermi and ATIC require an additional source of electrons and positrons
with an index $\g_a \approx 2$ at low energies and a cutoff $E_{\rm cut} \sim 500$ GeV.
However the presence of a bump at high energies in ATIC data contradicts
the smooth spectrum of Fermi and HESS.
The sources that produce featureless spectrum include the DM models with 
$M_{\rm DM} > 1$ TeV and several steps in DM annihilation process
as well as pulsars with an injection cutoff $E_{\rm inj.cut} \sim E_{\rm cut}$.
The sources that can produce ATIC bump include DM models with 
$M_{\rm DM} \sim E_{\rm cut}$ and few decay steps or DM models with 
$M_{\rm DM} \gg E_{\rm cut}$ and a significant contribution from a local 
clump.
Pulsars with $E_{\rm inj.cut} \gg E_{\rm cut}$ may give a sharp cutoff in the observed 
$e^+e^-$ flux and, possibly, step-like features due to a series of cooling break cutoffs from the 
youngest nearby pulsars.
This possibility is in tension with the Fermi data but consistent with the ATIC.

\bigskip
\bigskip

\noindent
{\large \bf Acknowledgments.}
\medskip

\noindent
The author is thankful to Mirko Boezio, Ilias Cholis, Joseph Gelfand, Lisa Goodenough, 
Michael Kuhlen, Neal Weiner, and Weiqun Zhang
for valuable discussions. This work is supported in part by the Russian Foundation
of Basic Research under grant RFBR 09-02-00253 and by the NSF grants
PHY-0245068 and PHY-0758032.



\bibliography{FApapers}         
\bibliographystyle{JHEP}

\end{document}